\documentclass[useAMS,usenatbib]{mn2e}

\title[The clustering of radio LRGs]{The clustering of radio galaxies at 
$z\simeq$ 0.55 from the 2SLAQ LRG survey}
\author[Wake et al.]{
\parbox{\textwidth}{
David A. Wake$^{1}$, 
Scott M. Croom$^2$,
Elaine M.\ Sadler$^2$, 
Helen M. Johnston$^2$
\vspace*{6pt} }\\
$^{1}$Department of Physics, University of Durham, South Road, Durham DH1 3LE, UK \\ 
$^{2}$School of Physics, University of Sydney, NSW 2006, Australia }

\begin{document}

\date{}

\pagerange{\pageref{firstpage}--\pageref{lastpage}} \pubyear{}

\maketitle

\label{firstpage}

\begin{abstract}
We examine the clustering properties of low--power radio galaxies 
at redshift $0.4<z<0.8$, using data from the 2SLAQ Luminous Red Galaxy 
(LRG) survey, and find that radio--detected LRGs (with typical 
optical luminosities of $3-5$\,L$_*$ and 1.4\,GHz radio powers in the 
range 10$^{24}$ to 10$^{26}$\,W\,Hz$^{-1}$) 
are significantly more clustered than a matched population of 
radio--quiet ($\la10^{24}$\,W\,Hz$^{-1}$) 
LRGs with the same distribution in optical luminosity and colour.  

The measured scale length of the two--point cross-correlation function
between the full LRG sample and the radio--detected LRGs is 
9.57$\pm$0.50\,$h^{-1}$\,Mpc, compared to  
8.47$\pm$0.27\,$h^{-1}$\,Mpc for the matched sample of radio--quiet LRGs; 
while the implied scale length of the auto-correlation function, $r_0$,  
is 12.3$\pm$1.2\,$h^{-1}$\,Mpc and 9.02$\pm$0.52\,$h^{-1}$\,Mpc for the 
radio--detected and radio--quiet samples respectively. 
We further interpret our clustering measurements in the halo model framework 
and demonstrate that the radio--detected LRGs have typical halo masses of 
10.1$\pm$1.4 $\times 10^{13} h^{-1}M_{\odot}$ and bias of 2.96$\pm$0.17, compared 
to 6.44$\pm$0.32 $\times 10^{13} h^{-1}M_{\odot}$ and 2.49$\pm$0.02 for the 
radio--quiet sample.  A model in which the radio--detected LRGs are almost 
all central galaxies within haloes provides the best fit to the measured 
clustering, and we estimate that at least $30\%$ of all 2SLAQ LRGs with the same 
clustering amplitude as the radio--detected LRGs are currently radio--loud. 

Our results imply that radio--detected galaxies in the 2SLAQ LRG sample 
typically occupy more massive haloes than other LRGS of the same optical
luminosity, so the probability of finding a radio--loud AGN in a massive 
galaxy at $z\sim0.55$ is influenced by the halo mass and/or cluster environment 
in addition to the well-known dependence on optical luminosity.  
If we model the radio--detected fraction of LRGs, 
F$_{\rm rad}$, as a function of halo mass M, 
then the data are well-fitted by a power law of the form 
$F_{\rm rad}\propto{\rm M}^{0.65\pm0.23}$. 
The observed relationship between radio emission and clustering strength 
could plausibly arise either through a higher fuelling rate of gas onto the 
central black holes of galaxies in the most massive haloes (producing more 
powerful radio jets) or through the presence of a denser IGM (which would 
provide a more efficient working surface for the jets, thus boosting their 
observed radio luminosity).  Further work is needed to determine which of 
these effects is dominant. 
\end{abstract}


\section{Introduction}
\subsection{Radio galaxies and their environment}
It has long been known that the hosts of powerful radio--loud AGN are 
massive early--type galaxies, and that the probability of such a galaxy 
hosting a radio source increases rapidly with optical luminosity and 
stellar mass (Auriemma et al.\ 1977; Best et al. 2005a; Mauch \& Sadler 2007).  
What remains less clear is the role (if any) of a galaxy's environment in 
determining whether it hosts a radio source. 

Studies carried out in the 1980s hinted at significant differences 
between the clustering properties of the powerful FR\,II\footnote{Fanaroff \& Riley (1974) 
divided radio galaxies into two classes based on their observed radio morphology.  
They found a correlation between morphology and radio luminosity, with 
less luminous (FR\,I) sources having a jet-like appearance and  
more luminous (FR\,II) sources having edge-brightened radio hotspots. } 
radio galaxies and the less powerful FR\,I sources.  
Based on an imaging study of 43 radio galaxies at $z<$ 0.3, Heckman et al.\ (1986) 
found that the most powerful radio sources (with radio luminosities above 
$10^{25.5}$\,W\,Hz$^{-1}$ at 408\,MHz) lay in regions of below--average galaxy 
density and often showed a disturbed morphology suggestive of a recent interaction 
with a gas-rich companion.  In contrast, less powerful radio sources appeared to be 
associated with morphologically--normal early-type galaxies in regions of high local 
galaxy density. Similar results were obtained by Prestage \& Peacock (1988), who used 
an angular cross--correlation technique to study the clustering environment of a sample 
of about 200 nearby ($z<$ 0.15) radio galaxies. They found that FR\,I radio galaxies 
lay in regions of significantly enhanced galaxy density, whereas the clustering 
environment of FR\,II sources was similar to that of the overall population of 
`normal' elliptical galaxies. 

The picture changed significantly in the mid 1990s with the work of Ledlow \& Owen (1996), 
who discovered that the division in radio power between FR\,I and FR\,II radio sources 
was a strong function of the optical luminosity of the host galaxy.  As a result, 
FR\,II radio sources are generally hosted by less optically luminous (and less massive) 
galaxies than FR\,I sources of similar radio power. Since more massive galaxies also 
tend to be more strongly clustered, this effect needs to be taken carefully into 
account when analysing the clustering properties of powerful radio sources. 
Ledlow \& Owen (1996) measured the bivariate radio luminosity function (RLF) 
of early--type galaxies in rich clusters, and found no statistically significant 
difference between the RLFs of galaxies in rich clusters and in the field.  
Their results suggested that the local environment plays little or no role 
in determining whether an early--type galaxy hosts a radio--loud AGN, and that 
the optical luminosity and other properties of the host galaxy are by far the most 
important parameters affecting radio source formation and evolution. 

\subsection{Measurements of radio--source clustering} 
The advent of large galaxy redshift surveys like the 2dFGRS and SDSS (Colless et al.\ 2001; 
York et al.\ 2000), combined with ``all--sky'' radio continuum surveys like NVSS and SUMSS 
(Condon et al.\ 1998; Bock et al.\ 1999) made it possible to assemble samples of thousands 
of objects with which to carry out statistical analyses of radio galaxies in the local 
universe (Best et al.\ 2005b; Mauch \& Sadler 2007).  The NVSS and SUMSS source catalogues 
are large and uniform enough that the characteristic imprint of large--scale structure can 
easily be seen in the angular correlation function (Blake \& Wall 2002; Blake et al.\ 2004).  
Convolving the angular clustering amplitude in these surveys with a characteristic redshift 
distribution N($z$) suggests that the present--day clustering length $r_0$ of radio galaxies 
is in the range of 7--10\,$h^{-1}$\,Mpc, corresponding to a clustering strength similar to 
optically--luminous elliptical galaxies in moderately rich environments (see e.g. Overzier 
et al.\ 2003). 

Recently, Best et al.\ (2007) have revisited the question of radio--source clustering 
using data sets much larger than those available to Ledlow \& Owen (1996). Using a sample 
of 625 nearby galaxy groups and clusters selected from the SDSS, they show that the brightest 
galaxies in groups and clusters (BCGs) are more likely to host a radio--loud AGN than other 
galaxies of the same stellar mass.  The probability is increased by up to a factor 
of two for the most massive galaxies (with stellar mass $\sim5\times10^{11}$\,M$_\odot$), 
and by over an order of magnitude for galaxies with stellar masses below $10^{11}$\,M$_\odot$. 
This enhanced likelihood of radio--loud AGN activity was only seen in the innermost 
regions of a group or cluster (i.e. within 0.2\,$r_{200}$, where $r_{200}$ is the Virial 
radius of the cluster). Best et al.\ (2007) argue that the radio properties of both BCGs 
and non-BCGs can be explained if the radio emission is mainly fuelled by cooling gas from 
an X--ray halo surrounding the galaxy. It therefore appears that although the radio 
properties of most galaxies in the local universe are largely unaffected by their 
environment, this is not true for  massive galaxies located in the central regions 
of clusters. 

\subsection{The 2SLAQ LRG radio sample at $z\sim0.55$}
At higher redshift, Sadler et al.\ (2007) recently combined data from the 2SLAQ Luminous 
Red Galaxy (LRG) redshift survey (Cannon et al.\ 2006) and the NVSS and FIRST radio surveys 
(Condon et al.\ 1998; Becker et al.\ 1995) to identify a volume--limited sample of 391 
radio galaxies at redshift $0.4<z<0.7$. They measured the redshift--space correlation 
between the radio--detected 2SLAQ LRGs and the full LRG sample, and found that the 2SLAQ 
radio galaxies were more strongly clustered than the overall 2SLAQ LRG population.  
Since the 2SLAQ radio galaxies as a class were also more optically luminous than the 
overall LRG sample it was unclear whether the increased clustering was a luminosity 
effect, or represented a genuine difference in the environments of radio--loud and 
radio--quiet LRGs at $z\sim0.55$. In the current paper, our goal is to answer this 
question by investigating the clustering properties of radio--loud 2SLAQ LRGs 
in more detail.  

\subsection{Radio--galaxy duty cycles}
The inferred lifetimes of the radio sources associated with massive galaxies (typically 
10$^6$--10$^8$\,yr; Parma et al.\ 1999) are significantly shorter than the ages of their 
parent galaxies, so it is generally assumed that all massive galaxies must cycle between 
radio-loud and radio-quiet phases over time.  Feedback mechanisms in which the hot 
intergalactic gas episodically cools to fuel a central AGN, and is then reheated by 
the ensuing radio jets (e.g. Binney \& Tabor 2005, Ciotti \& Ostriker 2007) provide 
a natural explanation for this process. 

For the 2SLAQ LRG sample, Johnston et al. (2008) have shown that the stellar 
populations of radio--detected and radio--quiet galaxies are generally indistinguishable. 
This is consistent with a picture in which `radio-mode' AGN feedback (Bower et al.\ 2006; 
Croton et al.\ 2006) regulates the star--formation rate in these massive galaxies, and all 
of them undergo radio--loud episodes when their central black hole is active and can power 
radio jets.  
If this is the case, then the fraction of galaxies which are detected as radio sources above 
some radio power P$_{\rm lim}$ simply represents the fraction of the radio--galaxy duty cycle 
for which a typical galaxy is a radio source at or above this level.

Observationally this is complicated by the fact that the radio luminosity function of AGN 
is very broad, spanning at least six orders of magnitude (Mauch \& Sadler 2007), and there 
is also a strong correlation between radio power and optical luminosity (Auriemma et al. 
1977, Best et al.\ 2005). The observed radio detection rate will therefore depend 
strongly on both radio power and galaxy luminosity.  In this paper we consider only the 
very luminous early--type galaxies which comprise the 2SLAQ LRG sample, so that the range 
in optical luminosity is small.  We also use the term `radio--detected' 
to refer to galaxies whose 1.4\,GHz flux density is higher than the 1--2\,mJy 
detection limit of the FIRST and NVSS radio surveys.  Since the 2SLAQ LRG radio sample 
is close to volume--limited (see Figure 7 of Sadler et al. 2007), this translates to 
a limiting radio luminosity of $\sim10^{24.2}$\,W\,Hz$^{-1}$. 

Throughout this paper, we assume a flat $\Lambda$--dominated cosmology with $\Omega_m$=0.27, 
$H_0$=70 km s$^{-1}$Mpc$^{-1}$, and $\sigma_8$=0.8 unless otherwise stated.

\section{The 2SLAQ LRG data samples}

\begin{figure*}
\vspace{6.5cm}
\includegraphics{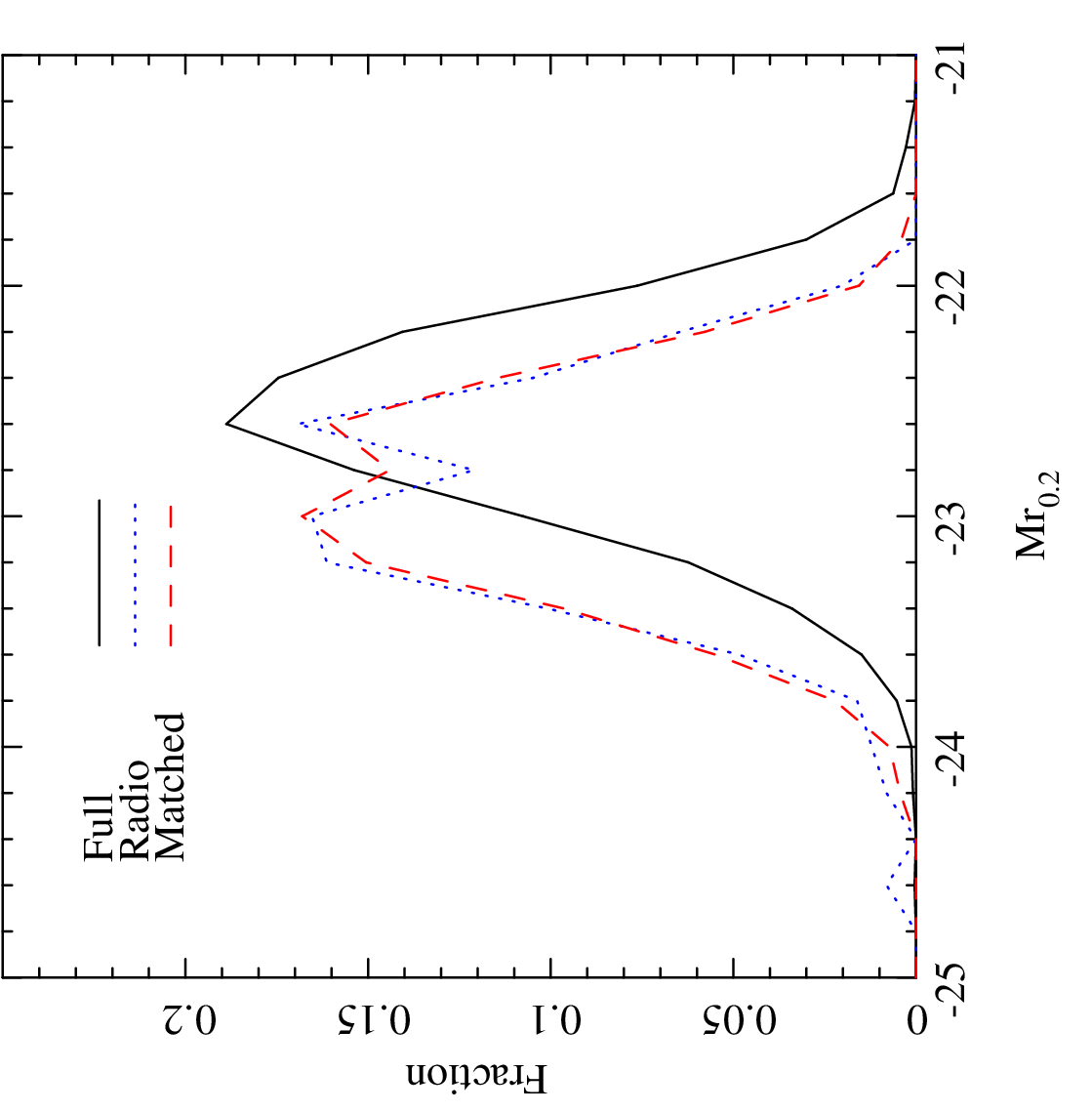}
\includegraphics{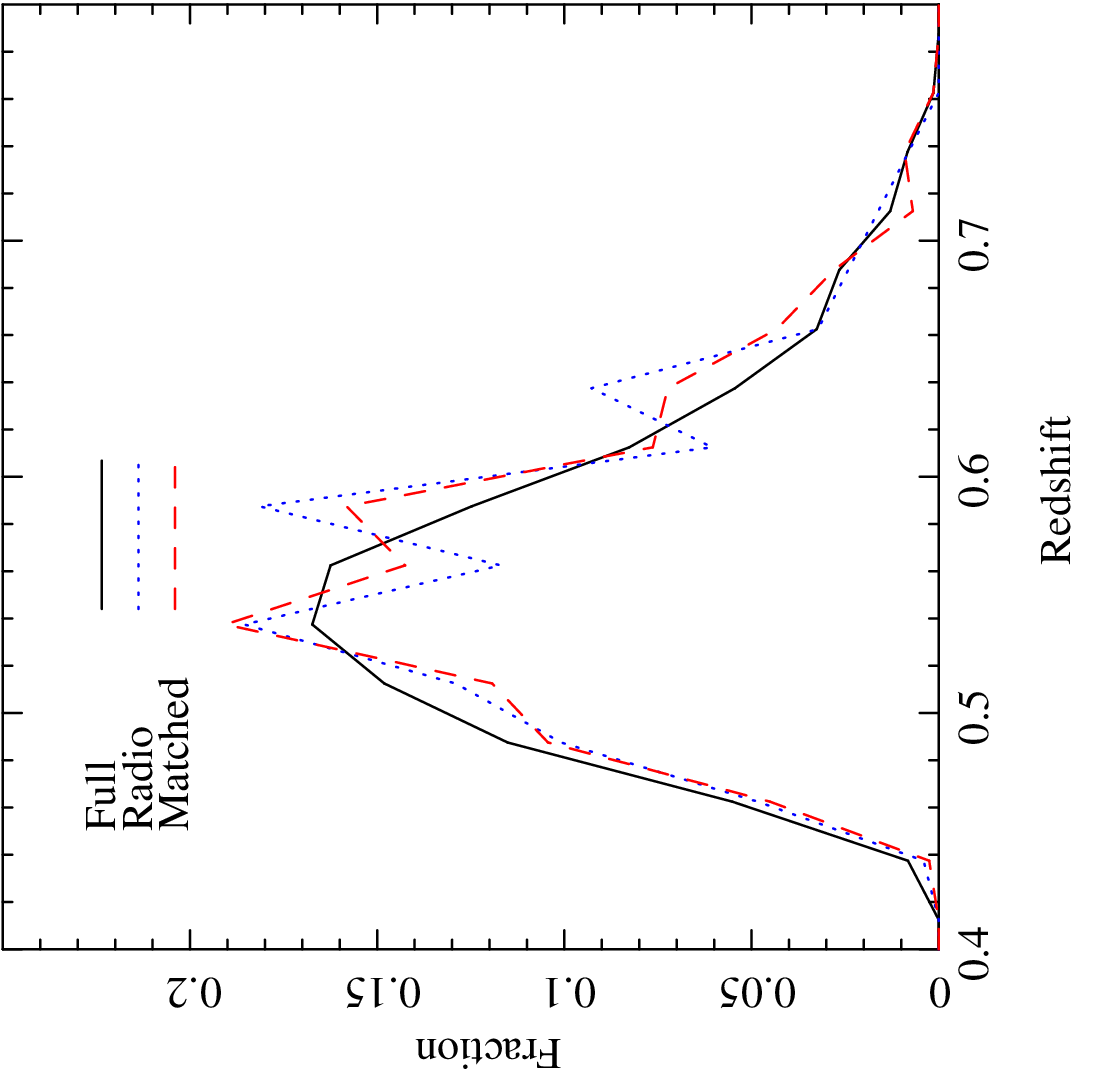}
\includegraphics{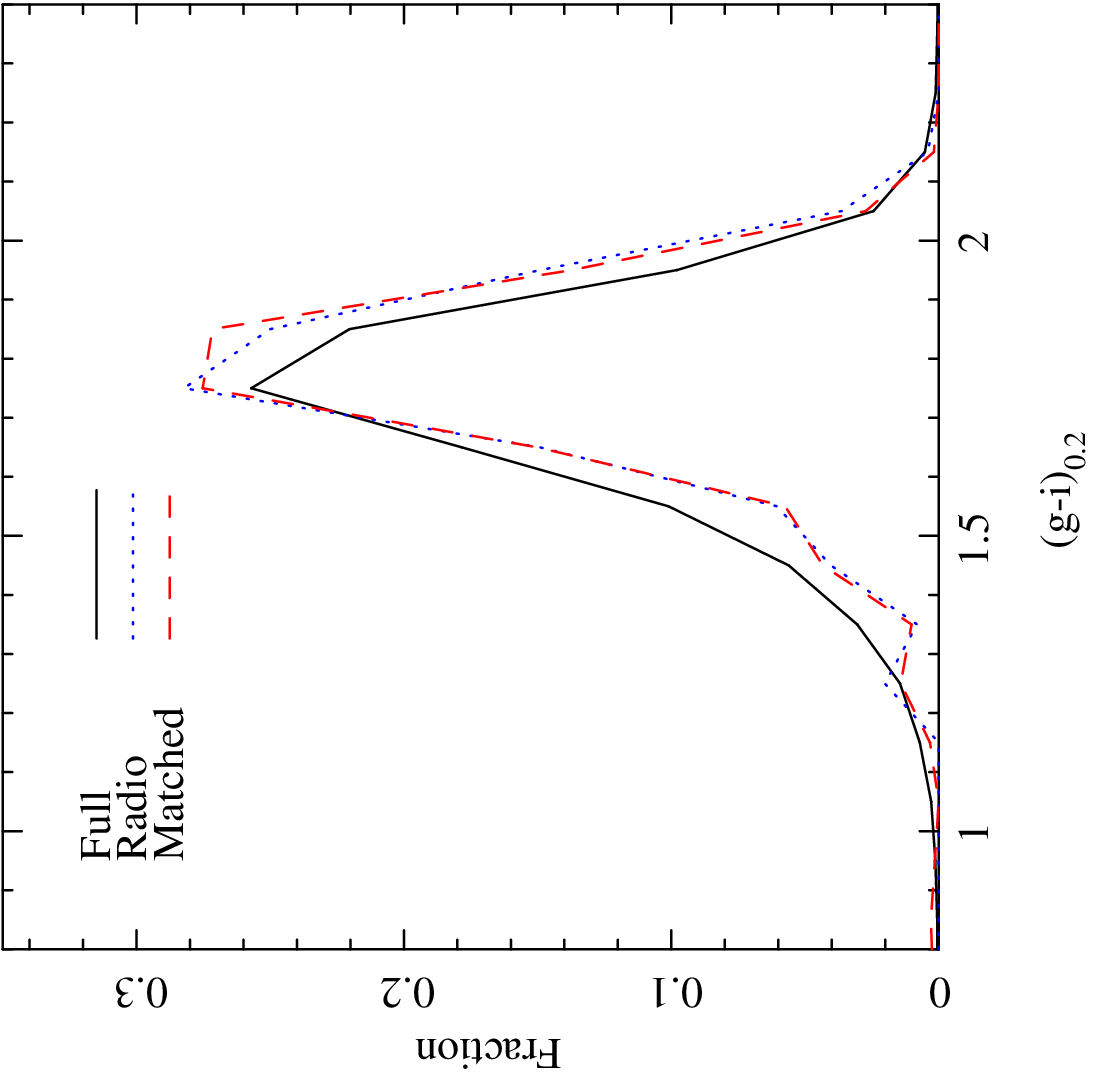}

\caption{\label{fig:raddist} The luminosity (left), redshift (middle) and colour (right) distributions 
of the full LRG sample (solid), the radio sample (dotted) and a sample randomly selected from the whole 
sample to match the luminosity, colour and z distributions of the radio sample (dashed).}
\end{figure*}

\begin{table}
  \begin{center}
    \caption{\label{tab:sample} Summary of the three 2SLAQ LRG samples used in this clustering study. }
    \begin{tabular}{c  l  r} 
      \multicolumn{1}{c}{Set} &
      \multicolumn{1}{c}{Properties } &
      \multicolumn{1}{c}{N} \\
      \hline \hline 
   1 & All `Sample 8' LRGs with good--quality $z$ & 7009 \\
   2 & Radio--detected LRGs from Set 1 & 250 \\
   3 & Luminosity--matched set of LRGs from Set 1 & 2750 \\ 
      \hline
    \end{tabular}
  \end{center}
\end{table}

In this paper we will consider the clustering of three data sets, as summarised in 
Table \ref{tab:sample}. All three samples are drawn from the full 2SLAQ LRG spectroscopic 
survey (Cannon et al.\ 2006).  The first data set consists of all main sample (`Sample 8') 
LRGs with high quality redshifts and $0.45<z<0.8$ (see Cannon et al.\ 2006 for details). 
The second set is a sub-sample of the first, and includes only the LRGs which have been 
detected as radio sources (see Sadler et al.\ 2007). Since we want to test whether  
radio--detected LRGs are more strongly clustered, we select a further sub-sample which 
has the same optical properties as the radio--detected sub-sample but does not contain 
any LRGs with detected radio emission. 

Figure \ref{fig:raddist} shows the distribution of the intrinsic luminosity, redshift and 
colour of the whole LRG sample and the radio--detected subsample, where we have used the 
K+e corrections as detailed in Wake et al. (2006) to generate the intrinsic luminosity and 
colours. This figure shows that the radio
sample is intrinsically more luminous, has a slightly higher typical redshift and is slightly 
redder. 
As a result of this, to generate a matched sample of LRGs without radio emission we must 
match for both optical luminosity and redshift.  To be complete we also match for colour, 
even though the difference in the distributions is marginal. We note that excluding this 
colour matching makes no difference to our results. 
We generate our sample by selecting the eleven LRGs from the whole sample that lie closest 
to each radio-detected LRG in colour-magnitude-redshift space. The final matched sample 
contains 2750 LRGs, which is the maximum that can be generated if we want to match the 
radio distributions without a large number of repeats ($\sim$15\%).

\section{The two--point cross--correlation function}
Since the space density of the radio LRG sample is very low, measurements of its auto-correlation 
function would be dominated by shot noise, particularly on small scales. In order to reduce this 
noise we cross-correlate the radio and matched samples with the full LRGs sample, which has a 
much higher space density.

The 2pt-cross-correlation function between two sets of objects {\it a} and {\it b}, $\xi_{ab}(r)$, 
is defined as a measurement of the excess probability above Poisson of finding an object {\it a} 
at a separation {\it r} from another object {\it b}. Here we wish to cross-correlate the radio 
and matched samples with the full LRG sample. We calculate this by comparing the number of pairs 
as a function of scale between the radio--detected (or matched) and full sample, with the number 
of pairs between the radio (or matched) and an unclustered (random) catalogue, which covers the 
same volume as the full sample such that

\begin{equation}
	\xi(s) = \frac{n_R}{n_f}\frac{N_{Rf}(r)}{N_{Rr}(r)} - 1,
\end{equation}
where $N_{Rf}$ and $N_{Rr}$ are radio-full and radio-random pair counts respectively, 
and $n_f$ and $n_r$ are the number of galaxies in the full and random samples. 

When making this calculation for our samples we must take into account the effect of the 
completeness varying across our survey. We follow the procedure described in detail by Wake 
et al.\ (2008), by up-weighting LRGs in areas of low completeness and using a random catalogue 
that has a constant space density over the angular mask of the survey. We exclude from our 
calculations regions that have $<$ 65\% completeness or that are close to bright stars.

We estimate the errors on our 2pt-cross-correlation function measurements using jack-knife 
re-sampling (Scranton et al.\ 2002; Zehavi et al.\ 2005). We split the 2SLAQ area into 74 equal 
area regions, minimising the noise on the covariance matrix whilst still removing regions larger 
than the scales we are interested in. We then repeatedly 
calculate each 2pt-function removing one area at a time to generate a full covariance matrix. 
Throughout we generate the pair counts using the KD-tree code in the NTROPY software package 
(Gardener et al.\ 2007).

The peculiar velocities of galaxies generate an error in the distance measurement to a galaxy 
along the line of sight, which results in distortions to $\xi$ known as redshift space distortions. 
To remove this effect we can calculate the clustering perpendicular ($r_p$) and parallel ($\pi$) 
to the line-of-sight ($\xi(r_p,\pi)$) and then integrate over the $\pi$ direction to 80 $h^{-1}$Mpc 
to give the 
projected correlation function ($w(r_p)$) such that  
\begin{equation}
	w_p(r_p) = 2\int^{\infty}_0 d\pi\xi(r_p,\pi).
\end{equation}
This can be expressed in terms of the real space correlation function $\xi(r)$ (Davis 
\& Peebles 1983) with

\begin{equation}
\label{eq:wrp}
	w_p(r_p) = 2\int^{\infty}_{r_p} rdr\xi(r)(r^2-r_p^2)^{-1/2}.
\end{equation}
If a power law of the form $\xi(r) = (r_0/r)^{-\gamma}$ is assumed then equation \ref{eq:wrp} 
can be solved analytically (Davis \& Peebles 1983).

\section{Clustering properties of radio--detected LRGs at $z\sim0.55$}
\label{sec:rad}

\begin{figure}
\vspace{7.5cm}
\includegraphics{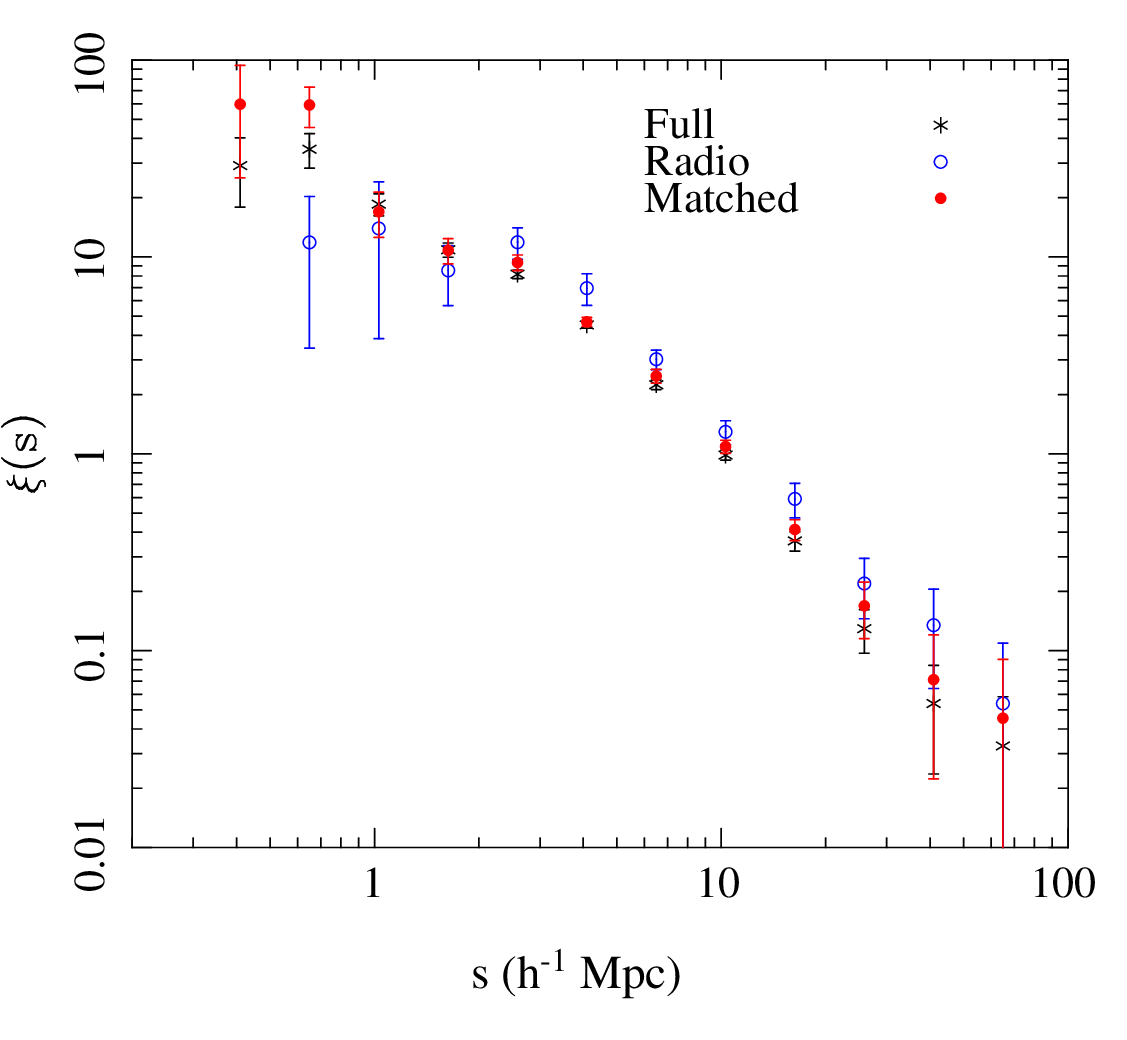}

\caption{\label{fig:xis} The redshift space 2pt-auto-correlation function for the full LRG 
sample (stars) and the redshift space 2pt-cross-correlation function for the radio sample 
(open circles) and a sample randomly selected from the whole sample to match the luminosity, 
colour and redshift distributions of the radio sample (filled circles).}
\end{figure}

\begin{figure}
\vspace{7.5cm}
\includegraphics{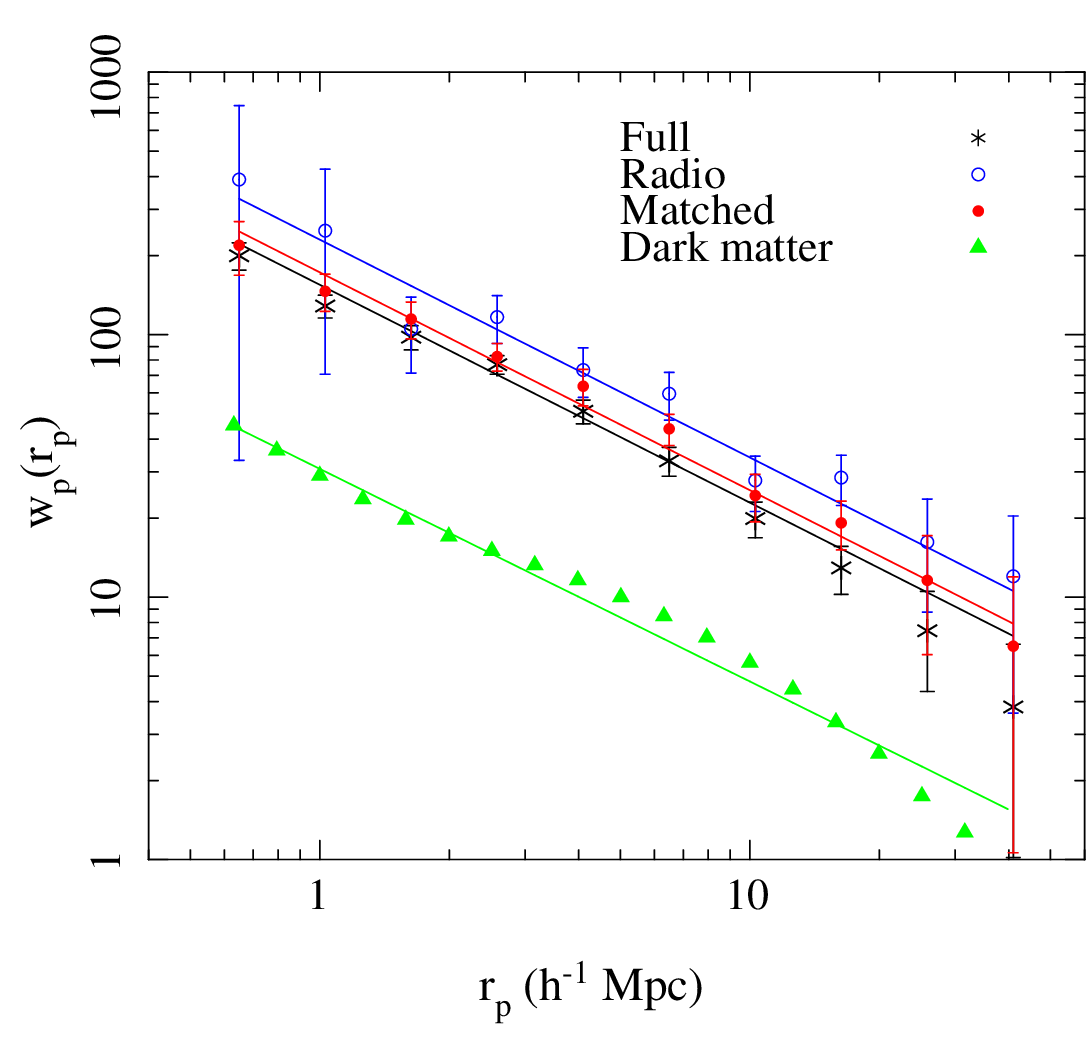}

\caption{\label{fig:wrp} The projected 2pt-auto-correlation function for the full LRG sample 
(stars) and dark matter (triangles), and the 2pt-cross-correlation function for the radio sample 
(open circles) and a sample randomly selected from the whole sample 
to match the luminosity, colour and redshift distributions of the radio sample (filled circles). 
The lines show power--law fits on scales $0.5 < r_p < 50 h^{-1}$ Mpc with slope, $\gamma$, 
fixed at 1.83 corresponding to the best fit value for the full 2SLAQ LRG sample.}
\end{figure}

\begin{figure}
\vspace{7.5cm}
\includegraphics{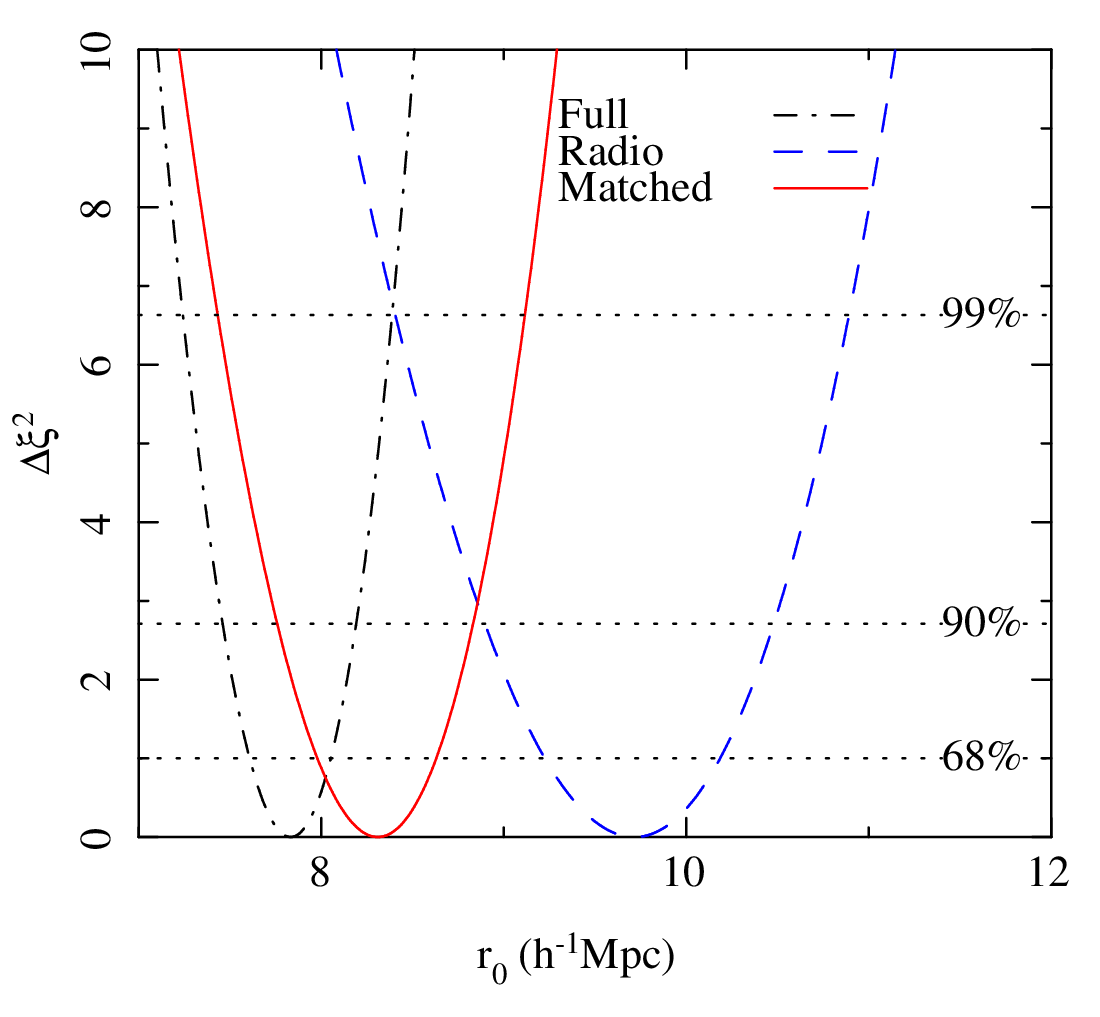}

\caption{\label{fig:fitdist} $\Delta\chi^{2}$ for the power law fits to the projected 
2pt-cross-correlation functions for the radio sample (blue dashed), a sample matching the luminosity, 
colour and redshift distributions of the radio sample (red solid), and the auto-correlation function 
of the full LRG sample (black dot-dashed). The slope, $\gamma$, is fixed at 1.83 corresponding to the best 
fit value for the full sample. The horizontal dotted lines show the 68\%, 90\% and 99\% confidence intervals.}
\end{figure}

\begin{table}
  \begin{center}
    \caption{\label{tab:fit} Values of the power-law fits to the projected 2pt-auto-correlation 
    function for the full sample and dark matter, and the 2pt-cross-correlation function for the 
    radio and matched 
    samples in the range $0.5 < r < 50 h^{-1}$ Mpc. Errors are at the 68\% confidence level.}
    \begin{tabular}{c  c  c  c  c} 
      \multicolumn{1}{c}{Sample} &
      \multicolumn{1}{c}{$r_0$ (h$^{-1}$\,Mpc)} &
      \multicolumn{1}{c}{$\gamma$}&
      \multicolumn{1}{c}{$\chi_{red}^2$}&
      \multicolumn{1}{c}{$r_0$ (fixed $\gamma$)}\\
      \hline \hline 
	Full & $7.66^{+0.16}_{-0.17}$ & $1.83^{+0.04}_{-0.04}$ & 1.6 &$7.66^{+0.16}_{-0.17}$ \\
	Matched& $8.47^{+0.27}_{-0.27}$ & $1.78^{+0.05}_{-0.05}$ & 1.4  & $8.31^{+0.22}_{-0.23}$ \\
	Radio& $9.57^{+0.51}_{-0.50}$ & $1.75^{+0.10}_{-0.10}$ & 1.8 & $9.72^{+0.49}_{-0.46}$   \\
	Dark Matter& $3.25^{+0.02}_{-0.02}$ & $1.81^{+0.02}_{-0.02}$ & --  & --    \\
      \hline
    \end{tabular}
  \end{center}
\end{table}

Figures \ref{fig:xis} and \ref{fig:wrp} shows the 2pt-auto-correlation function for the full 
sample, and the 2pt-cross-correlation function for the radio--detected and matched samples in redshift 
space and projection respectively. In order to compare the clustering strengths of these samples 
we make $\chi^2$ fits to $w(r_p)$ assuming a power law form for $\xi(r)$ of $(r/r_0)^{-\gamma}$ 
using the analytical solution to equation \ref{eq:wrp}. We use the full covariance matrices generated 
using the jack-knife resampling technique and fit over $0.5 < r_p < 50 h^{-1}$ Mpc. 
Values of $r_0$ and $\gamma$ for these fits are given in Table \ref{tab:fit}. 
Since the values of the slope, $\gamma$, are consistent between the three samples we refit using 
the $\gamma$ from the fit to the full LRG sample. The values of $r_0$ from these fits are also given 
in Table \ref{tab:fit} and the $\Delta\chi^2$ distributions are shown in Figure \ref{fig:fitdist}.

Figure \ref{fig:fitdist} and Table \ref{tab:fit} show that the radio--detected LRGs are more 
clustered (with a significance of 97\%) than other LRGs with similar optical luminosity.
When considering Figures 2, 3 and 4 and the fits given in Table
\ref{tab:fit}, it is important to remember that the amplitude of the cross-correlation of the radio
or luminosity matched samples will be lower than the auto-correlation for these samples,  
since they are cross-correlated with the full LRG sample which has a lower clustering strength.
On large ($>$ 1 Mpc) scales, where the clustering is determined by pairs of galaxies 
in separate dark matter haloes, one would expect the cross-correlation function to be 
the geometric mean of the auto-correlation of the two samples. On smaller scales where 
the clustering is dominated by pairs within haloes, the cross-correlation function will 
depend on the relative distribution of the two samples within haloes and would only
be the mean if the distribution was identical. We can therefore use the cross-correlation 
on large scales to calculate the auto-correlation function, where

\begin{equation}
\label{eq:xic}
	\xi_2 = \xi_{12}^2 / \xi_1.
\end{equation}
This gives values of $r_0$ for the auto-correlation function of 12.3 $\pm$ 1.2 and 
9.02 $\pm$ 0.52 $h^{-1}$\,Mpc for the radio--detected and matched samples respectively. 

We can estimate the large scale bias ($b$) for our three populations as 
$b = (w_{pGal}/w_{pDM})^{1/2}$ where $w_{pDM}$ is generated using the linear theory 
power spectrum as described in section \ref{sec:halo} and the projected auto-correlation function for the matched and radio--detected 
samples is calculated using Equation 4.  We define the large scale bias here as the 
weighted mean of $(w_{pGal}/w_{pDM})^{1/2}$ within $2 < r < 50\,h^{-1}$\,Mpc. 
We can then use the dependence between the dark matter halo mass and bias estimated 
from the halo mass function (Sheth \& Tormen 1999) to relate this bias 
to a typical halo mass. Table \ref{tab:fitauto} gives the mean bias and halo mass 
for each of the three samples.

\begin{table}
  \begin{center}
    \caption{\label{tab:fitauto} The scale length of 2pt-auto-correlation function ($r_0$) for the 
    three samples along with the inferred large scale bias and typical dark matter halo mass. 
    Errors are at the 68\% confidence level.}
    \begin{tabular}{c  c  c  c} 
      \multicolumn{1}{c}{Sample} &
      \multicolumn{1}{c}{$r_0$ ($h^{-1}$ Mpc)} &
      \multicolumn{1}{c}{bias}&
      \multicolumn{1}{c}{M$_{\rm DH}(10^{13}h^{-1}M_{\sun})$}\\
      \hline \hline 
	Full & 7.66$\pm$0.17 & 2.14$\pm$0.05 & 3.1$\pm$0.2 \\
	Matched& 9.02$\pm$0.52 & 2.52$\pm$0.16 & 5.7$\pm$1.1 \\
	Radio& 12.3$\pm$1.2 & 3.11$\pm$0.29 & 10.6$\pm$1.9 \\
      \hline
    \end{tabular}
  \end{center}
\end{table}

Our finding that the radio--detected 2SLAQ luminous red galaxies (which have typical 1.4\,GHz 
radio powers of 10$^{24}$ to 10$^{26}$\,W\,Hz$^{-1}$) are significantly more strongly-clustered 
than LRGs of similar optical luminosity which are not detected as radio sources (and so are weaker 
than $\sim10^{24}$\,W\,Hz$^{-1}$) implies that the radio--galaxy duty cycle $z\sim0.55$ is affected 
by at least one factor which is linked to the clustering environment, and we investigate this 
further in the next section. 

\section{Halo Models}
\label{sec:halo}

\begin{table*}
  \begin{center}
    \caption{\label{tab:fithod} Details of the HOD fits to the auto-correlation function 
    for the full sample and the cross-correlation function for the matched and radio samples. }
    \begin{tabular}{c  c  c  c  c  c  c  c  c}
      \multicolumn{1}{c}{Selection} &
      \multicolumn{1}{c}{Density} &
      \multicolumn{1}{c}{M$_{\rm min}$} &
      \multicolumn{1}{c}{M$_1$} &
      \multicolumn{1}{c}{$\alpha$} &
      \multicolumn{1}{c}{$\chi_{red}^2$}&
      \multicolumn{1}{c}{$b_{g}$}&
      \multicolumn{1}{c}{M$_{\rm eff}$}&
      \multicolumn{1}{c}{F$_{\rm sat}$}\\
      \multicolumn{1}{c}{} &
      \multicolumn{1}{c}{($10^{-4}h^{3}Mpc^{-3}$)} &
      \multicolumn{1}{c}{($10^{13}h^{-1}M_{\sun}$)} &
      \multicolumn{1}{c}{($10^{13}h^{-1}M_{\sun}$)} &
      \multicolumn{1}{c}{} &
      \multicolumn{1}{c}{}&
      \multicolumn{1}{c}{}&
      \multicolumn{1}{c}{($10^{13}h^{-1}M_{\sun}$)}&
      \multicolumn{1}{c}{(\%)}\\
      \hline \hline
        Full  & 1.55 $\pm$ 0.02 & 1.96 $\pm$ 0.02 & 24.6 $\pm$ 1.1 & 1.93 $\pm$ 0.13 & 1.2 & 2.15 $\pm$ 0.01 & 4.56 $\pm$ 0.09& 5.2 $\pm$ 0.7\\
        Matched  & 0.45 $\pm$ 0.01 & 4.70 $\pm$ 0.09 & $ >59.5$ & $>0.85$ & 1.0 & 2.49 $\pm$ 0.02 & 6.44 $\pm$ 0.32 & 0.9$\pm$1.9\\
        Radio & 0.15 $\pm$ 0.05 & 9.65 $\pm$ 2.2 &  $>80.0$ & $>0.6$  & 0.82 & 2.96 $\pm$ 0.17 & 10.1 $\pm$ 1.4 & 0.0 $\pm$ 4\\
      \hline
    \end{tabular}
  \end{center}
\end{table*}

\begin{figure*}

\vspace{8.0cm}
\includegraphics{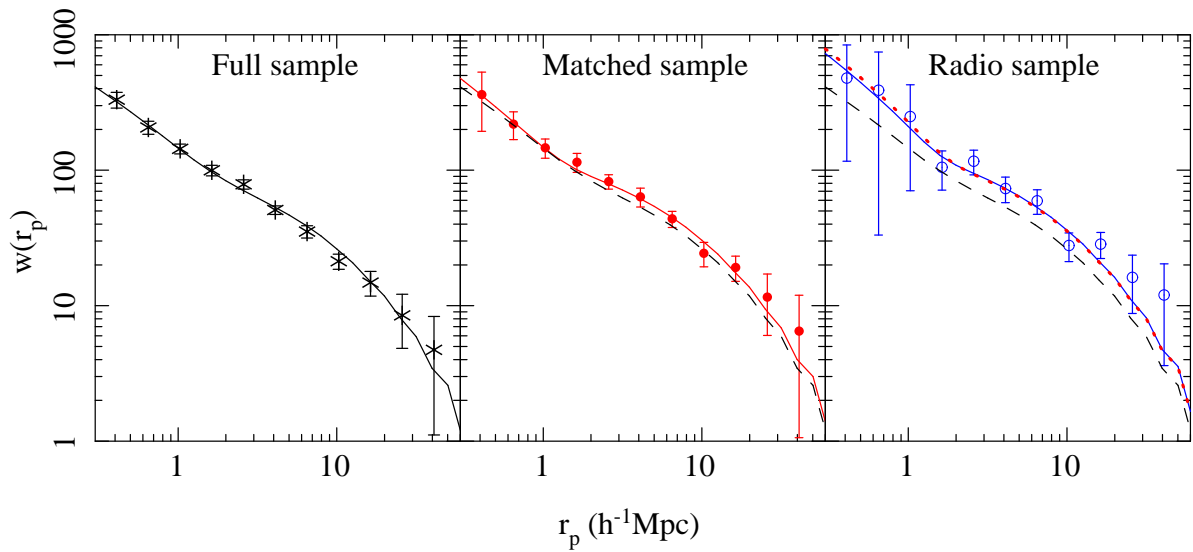}

\caption{\label{fig:HODfit} The projected 2pt-auto-correlation function for the full LRG sample 
(stars) and the projected 2pt-cross-correlation function for the radio sample (open circles) and 
a sample randomly selected from the whole sample to match the luminosity, colour and z distributions 
of the radio sample (filled circles). The solid lines show the best fitting HODs on scales 
$0.32 < r_p < 50 h^{-1}$ Mpc. The dashed lines show the fit to the full sample, and the dotted line 
the fit achieved applying a power-law radio fraction to the matched HOD fit (see text for details).}
\end{figure*}

\subsection{Model parameters} 
The halo model assumes that the galaxy clustering signal encodes information about the Halo Occupation 
Distribution (HOD;how the galaxies populate dark matter haloes), in particular how the HOD depends on h
alo mass (see e.g. Jing et al 1998, Ma \& Fry 2000, Peacock \& Smith 2000, Seljak 2000, Scoccimarro et al. 
2001, Berlind \& Weinberg 2002). We have successfully applied this technique to the 2pt-correlation 
function of the 2SLAQ LRGs 
(Wake et al.\ 2008), and use the same techniques here to gain a further understanding of how the radio 
galaxies are distributed within dark matter haloes. We give below a brief outline of our halo model 
description of the clustering, and refer the reader to Wake et al.\ (2008) for further details. 

In the halo model, every galaxy is associated with a halo and all haloes are 200 times the background 
density whatever their mass $M$.  Sufficiently massive haloes typically host more than one galaxy.  
The halo model we use distinguishes between the central galaxy in a halo and the others, which are 
usually called satellites.  

The fraction of haloes of mass $M$ which host centrals is modelled as 
\begin{equation}
\label{eq:Ncen}
	\langle N_c|M\rangle = \exp(-M_{min}/M). 
\end{equation}
Only haloes which host a central may host satellites.  In such haloes, the number of satellites is drawn 
from a Poisson distribution with mean 
\begin{equation}
 \label{eq:Nsat}
	\langle N_s|M \rangle = (M/M_1)^{\alpha}.
\end{equation}
Thus, the mean number of galaxies in haloes of mass $M$ is 
\begin{equation}
 \label{eq:Ntot}
 \langle N|M\rangle = \langle N_c|M\rangle[1 + \langle N_s|M \rangle], 
\end{equation} 
and the predicted number density of galaxies is 
\begin{equation}
\label{eq:den}
	n_g =  \int dM\, n(M)\, \langle N|M\rangle,
\end{equation}
where $n(M)$ is the halo mass function, for which we use the parametrisation given by Sheth \& Tormen (1999).

We further assume that the satellite galaxies in a halo trace an NFW profile (Navarro et al. 1996) 
around the halo centre, and that the haloes are biased tracers of the dark matter distribution.  
The halo bias depends on halo mass in a way that can be estimated directly from the halo mass function 
(Sheth \& Tormen 1999).  With these assumptions, the halo model for $\xi(r)$ is completely specified 
(e.g. Cooray \& Sheth 2002). Our model for the real space 2pt-auto-correlation function is given in 
detail in Wake et al. (2008), where the mean number density of central-satellite pairs within haloes 
of mass $M$ is $n(M)\,\langle N_c|M\rangle\,\langle N_s|M \rangle$, and the mean number density of 
distinct satellite-satellite pairs is 
$n(M)\,\langle N_c|M\rangle\,\langle N_s|M \rangle^2/2$. 
In this work we use the 2pt-cross-correlation function, where we are cross-correlating a sub-sample 
of LRGs (either the radio--detected or matched samples) with the full sample of LRGs. We model the 
cross-correlation function in the halo model framework analogously to the auto-correlation function 
with the mean number density of central-satellite pairs, 
\begin{equation}
\label{eq:ncs}
n_{cs}(M) = n(M)\,\langle N_{c1}|M\rangle\,[\langle N_{s}|M \rangle + \,\langle N_{s1}|M \rangle],
\end{equation}
 and the mean number density of distinct satellite-satellite pairs, 
\begin{equation}
\label{eq:nss}
n_{ss}(M) =  n(M)\,\langle N_{c1}|M\rangle\,\langle N_{s}|M \rangle\,\langle N_{s1}|M \rangle,
\end{equation}
where the pairs are between the full sample and sub-sample, and the terms with a subscript 1 are 
for the sub-sample and those without are for the full sample. 

Since in our definition of the HOD a halo must contain a central if it is to contain a satellite, 
whether for the full sample or sub-sample, then only haloes that have a central in the sub-sample 
will contribute to the cross-correlation function. This is why only the fraction of haloes which 
contain a central of the sub-sample $\langle N_{c1}|M\rangle$ appears in equations \ref{eq:ncs} 
and \ref{eq:nss}, not the equivalent fraction for the full sample $\langle N_{c}|M\rangle$.

Our model for the real-space 2pt-cross-correlation function is then 
\begin{equation}
 \xi_{12}(r) = 1+\xi_{cs}(r) + 1+\xi_{ss}(r) + \xi_{2h12}(r)
\end{equation}
where 
\begin{eqnarray}
1+\xi_{cs12}(r) &=& \int dM\, {n_{cs12}(M)\over n_{g1}n_{g2}}\, {\rho(r|M)\over M} \\
1+\xi_{ss12}(r) &=& \int dM\, {n_{ss12}(M)\over n_{g1}n_{g2}}\, {\lambda(r|M)\over M^2} 
\end{eqnarray}
and 
\begin{equation}
\xi_{2h12}(r) = \int \frac{dk}{k}\,\frac{k^3P_{2h12}(k)}{2\pi^2}\,\frac{sinkr}{kr}
\end{equation}
with 
\begin{eqnarray}
 P_{2h12}(k) &=& b_{g1}(k)b_{g2}(k)\,P_{\rm Lin}(k),
\end{eqnarray}
where
\begin{eqnarray}
  b_g(k) &=& \int dM {n(M)\over n_g}\, b(M)\,
 \Bigl[\langle N_c|M\rangle + \langle N_s|M\rangle u(k|M)\Bigr].\nonumber
\end{eqnarray}
In the expressions above,
 $\rho(r|M)$ is the density profile of haloes of mass $M$,
 $\lambda(r|M)$ denotes the convolution of two such profiles, 
 $u(k|M)$ is the Fourier transform of $\rho(r|M)/M$, 
 and $P_{\rm Lin}(k)$ denotes the linear theory power spectrum.  
In practise, we approximate $b_g(k)$ by its value $b_{g}$ 
at $k=0$ (equation~\ref{eq:blin}).  
All these quantities, along with the mass function $n(m)$ and halo bias 
factor $b(M)$, are to be evaluated at the redshift of interest.
We then calculate $w(r_p)$ from $\xi(r)$ using equation~(\ref{eq:wrp}).

\subsection{Application to the 2SLAQ LRG data} 
This model for the cross-correlation-function makes the assumption that the two galaxy samples 
occupy the same haloes, and that the satellite galaxies of both samples follow the same profile 
within the haloes. Whilst this is not necessarily valid for two independent galaxy samples, it 
should hold here as one sample is always a subset of the other.

This form of the halo model makes the explicit assumption of a volume--limited sample of galaxies. 
The 2SLAQ LRG sample as a whole is magnitude--limited rather than  volume--limited, 
but (as may be seen from Table 4 of Sadler et al.\ 2007), the radio--detected LRGs are significantly 
brighter than the LRG sample as a whole.  For this reason, Sadler et al.\ (2007) argue that the 
sample of radio--detected 2SLAQ LRGs is close to volume-limited, with no strong correlation 
between absolute magnitude (or radio luminosity) and redshift. 

Nevertheless, since the use of a magnitude--limited sample may have the effect of biasing any derived 
quantities we may wish to measure from the HOD (such as the typical halo mass or satellite fraction)  
we carried out some further investigations.  In the Appendix, we investigate the consequences 
of applying the halo model to magnitude--limited galaxy samples using the latest semi-analytic galaxy 
formation models (Font et al. 2008) applied to the Millennium simulation (Springel et al. 2005). 
We show that the halo model successfully recovers the effective halo mass to within 5\% and so 
we are confident in applying this model to the 2SLAQ LRG samples.

\begin{figure}
\vspace{10.0cm}
\includegraphics{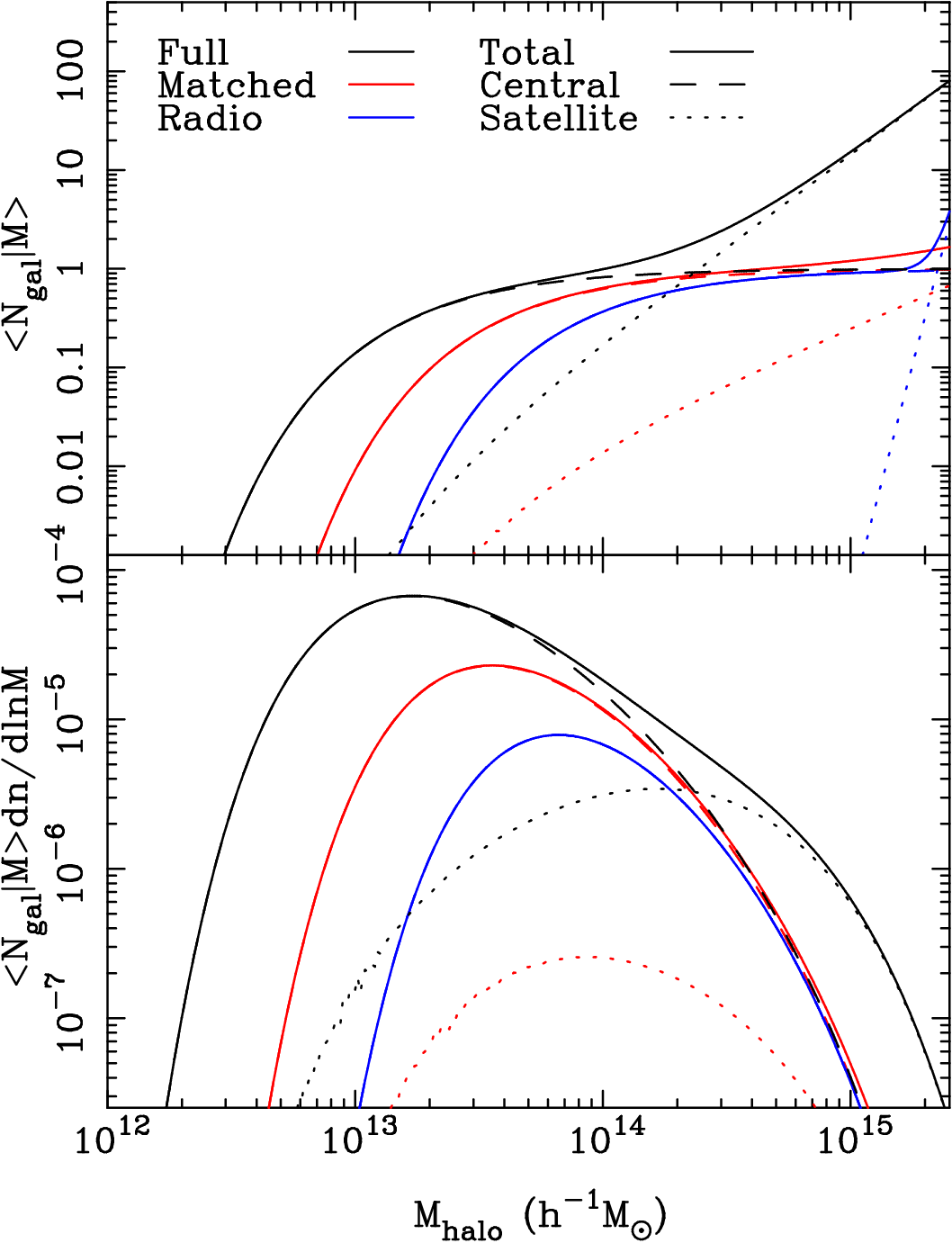}
\caption{\label{fig:NM} The mean number of LRGs per halo as a function of halo mass (top) 
and the mean number of LRGs per halo normalised by the number of haloes as a function of mass 
(bottom) for the best fitting HODs of the three samples. The full sample is shown by the black 
lines, the radio sample is shown by the blue lines and the luminosity, colour and redshift 
matched sample is shown by the red line. The total, central and satellite contributions are 
shown by the solid, dashed and dotted lines respectively. Whilst the spike at large halo masses 
in the radio HOD is the best fit, the HOD parameters describing the satellites are very poorly 
fit and a model with no satellites represents a better fit to the data.}
\end{figure}

We can now fit this model to the measured clustering and find the best fitting HOD parameters 
$M_{\rm min}$, $M_1$, and $\alpha$. We first make a $\chi^2$ fit to the full LRG sample, 
fitting both the measured 2pt-auto-correlation function and the measured space density of
$1.55\times10^{-4}h^3Mpc^{-3}$ as was done by Wake et al. (2008). 
We then use the resulting best fit HOD when fitting the cross-correlation function of the 
matched and radio samples.  We note that including the density for the radio sample in this 
fitting process produces a poor fit to the clustering. This is not surprising, since only
a small fraction (3--4\%) of the 2SLAQ LRGs are detected as radio sources and so their 
overall space density is very low. Our HOD parametrisation assumes that all haloes above 
a certain mass contain at least one galaxy. Therefore the lower the galaxy density, the 
higher the typical mass and the higher the clustering. 
This is a good parametrisation for the whole galaxy population limited by 
optical luminosity, due to the correlation between central galaxy stellar mass and 
the halo mass. However, if the duty cycle of radio emission is such that only a 
fraction of LRGs that could be radio loud actually show emission at a given time, 
then the measured space density of the radio--detected objects is a product of both the 
parent population space density and the fractional duty cycle. 

To account for this, we could modify our HOD to include a term determining the fraction 
of LRGs which are currently radio--emitting. Initially we chose a simpler approach by 
not including the density and fitting the clustering alone. 
We can then compare the density predicted by the best--fitting HOD to the measured value 
and determine the fraction of LRGs with the same clustering as the radio sources which 
are currently radio--loud. 
By doing this, we are assuming that the relationship between halo mass and the probability 
of a galaxy having detectable radio emission takes the same form that we use to relate 
halo mass and luminosity, multiplied by some fraction which is independent of halo mass.   

\subsection{Results} 
Figure \ref{fig:HODfit} shows the best fitting HODs for the three samples with the best--fitting 
parameters given in Table \ref{tab:fithod}. We can use the HODs to calculate some other useful 
quantities such as the average linear bias ($b_g$), the effective halo mass ($M_{\rm eff}$) and 
the satellite fraction ($F_{\rm sat}$), where
\begin{equation}
\label{eq:blin}
	b_{g} =  \int dM n(M) b(M)<N>/n_g,
\end{equation}
\begin{equation}
\label{eq:Meff}
	M_{eff} =  \int dM M n(M) <N>/n_g,
\end{equation}
and
\begin{equation}
\label{eq:Fsat}
	F_{sat} =  \frac{\int dM n(M) <N_s|M>}{\int dM n(M) <N|M>}.
\end{equation}
The values of these quantities for each sample are given in
Table \ref{tab:fithod}.

As we would expect from the relative clustering strength, 
we find a significantly higher minimum halo mass ($M_{\rm min}$)
for the radio sample than the matched sample, confirming
that the radio galaxies are typically found in more massive
haloes.  This is further confirmed by the calculated values 
of $M_{\rm eff}$ and $b_g$. 

The parameters $M_1$ and $\alpha$ describing the satellite
population are quite poorly constrained by these fits, with a large 
range of parameter space with high $M_1$ or $\alpha$ having
a very similar $\chi^2$. This is because these samples have very
low numbers of satellites, with the best fitting HODs for
the matched and radio samples being consistent with having
no satellite galaxies.

The best--fitting HODs are plotted in Figure \ref{fig:NM}, 
which shows the number of galaxies as a function of halo mass (top) 
and the number of galaxies weighted by the number of haloes as a 
function of halo mass (bottom).  Each plot shows the total galaxy 
distributions, as well as those of the central and satellite galaxies 
independently.  These plots clearly show that for both the radio--detected 
and matched samples, satellite galaxies are only present in the most massive haloes. 
Since there are so few haloes with these high masses, their total contribution 
is very low. This also explains why M$_1$ and $\alpha$ are given 
as 1$\sigma$ lower limits in Table \ref{tab:fithod}. 
When M$_1$ becomes significantly large there is essentially 
no contribution to the clustering from satellites since there are so few 
haloes at these masses, so the fit is equally as good for any large value 
of M$_1$ or $\alpha$. In fact, fitting a model with no satellite galaxies 
produces a $\chi^2$ only marginally higher for both the radio--detected and 
matched samples, and so gives significantly better fits overall because of 
the larger number of degrees of freedom in the model with satellites.

\subsection{HOD models and the radio--galaxy duty cycle }

We derive a space density $0.15 \pm 0.05 \times10^{-4}h^3$\,Mpc$^{-3}$ for the 
best--fitting HOD to the radio--detected LRG sample.   This can now be compared 
to the measured density of 2SLAQ LRGs which are currently detected as radio 
sources\footnote{This value is found   
simply by dividing the total number of radio--detected LRGs by the survey 
completeness and co-moving volume between $0.44<z<0.76$ in the 2SLAQ survey area. 
Our clustering analysis uses the full 2SLAQ LRG sample, which is magnitude--limited 
and so includes a smaller contribution from lower--luminosity galaxies than would be 
expected in a volume--limited sample.  For this reason, we argue that the simple space 
density derived above is the appropriate one to use in this comparison. If we integrate 
over the radio luminosity function (RLF) in Table 5 of Sadler et al.\ (2007), which uses 
the V$_{\rm max}$ estimator to correct for the effects of the magnitude limit, we obtain  
a slightly higher value of $0.067 \pm 0.004 \times10^{-4}h^3$\,Mpc$^{-3}$ for the volume--limited 
sample, which would    
increase the estimated radio--loud fraction from $\sim30$\% to $\sim45$\%}, 
$0.041 \pm 0.003 \times10^{-4}h^3$\,Mpc$^{-3}$.   
Taking the ratio of these two numbers suggests that about 30\% of all 2SLAQ LRGs with 
the same clustering properties as the radio--detected sample are currently radio--loud.  
This in turn implies a fairly high duty cycle for radio--loud AGN in the central galaxies 
of clusters at $z\simeq$0.55, and is consistent with the radio detection rate of the 
brightest 2SLAQ LRGs (28$\pm$8\% for M$_{0.2,r}<-23.75$ from Table 4 of Sadler et al.\ 2007).  

The above discussion relies on our assumption that the form of the HOD for the 
radio--detected LRGs is the same as the standard form for a luminosity--limited 
sample. In that case the only mass--dependent probability of an LRG being radio--loud 
is determined by $M_{min}$, and there is then a halo mass--independent probability of 
$\sim$30\% of a galaxy being currently powerful enough to be detected as a radio source 
at $\ga10^{24}$\,W\,Hz$^{-1}$.  
However it may be the case that the HOD for radio LRGs takes a different form from 
that given in equations \ref{eq:ncs} and \ref{eq:nss}, since one might  
expect it to consist of the standard form plus some term describing how the fraction of 
radio loud LRGs depends on halo mass ($F_{\rm rad}(M)$). 

We can attempt to determine $F_{\rm rad}(M)$ by trying to modify the best fitting HOD to the 
matched sample in such away that we reproduce the clustering and space density of the 
radio--detected LRG population. We chose to model $F_{\rm rad}(M)$ as a power law such that

\begin{equation}
 \label{eq:Frad}
	F_{\rm rad}(M) = f_r(M/M_{\rm rad})^{\beta},
\end{equation}
 where $M_{\rm rad} = 10^{14}h^{-1}M_{\sun}$ and $F_{\rm rad}(M) \leq$ 1.

\begin{figure}
\vspace{11.0cm}
\includegraphics{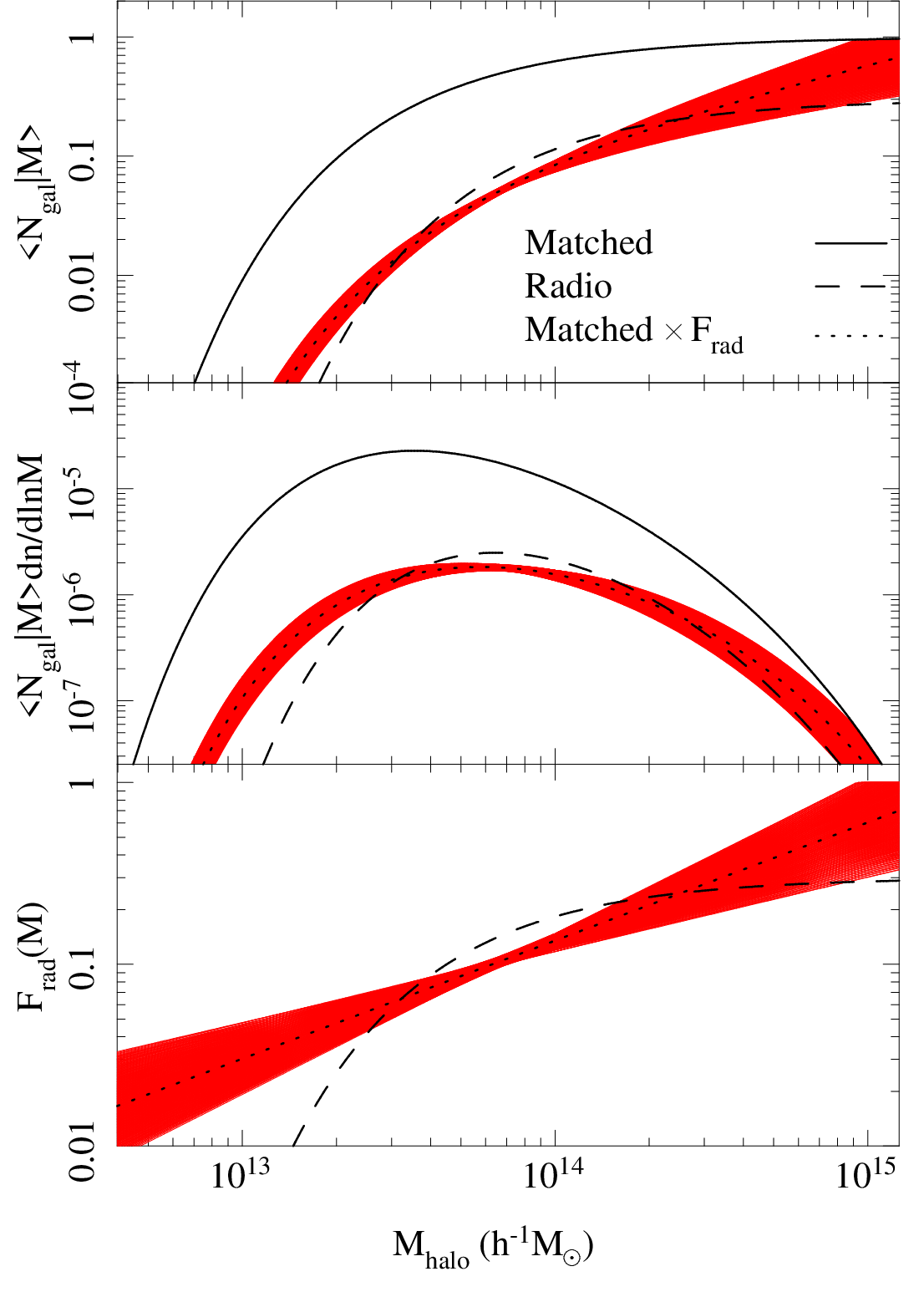}
\caption{\label{fig:NMfrac} The mean number of LRGs per halo (top), the mean number of 
LRGs per halo normalised by the number of haloes (middle), the fraction of radio--loud 
LRGs (bottom), all as a function of halo mass. The solid line shows the best fit to 
the matched sample, the dashed line shows the best fit to the radio sample when the 
density is matched, and the dotted line shows the best fitting matched HOD plus 
power-law radio fraction to the radio clustering. The red area shows the 1$\sigma$ 
error on the dotted line.}
\end{figure}

Figure \ref{fig:NMfrac} shows the best fitting power law for $F_{\rm rad}(M)$, HOD and 
space density distribution, along with the best fitting standard HOD given in Table 
\ref{tab:fithod}. The fit to the clustering is shown as the dotted line in Figure 
\ref{fig:HODfit}. The best fit has a slope $\beta = 0.65 \pm 0.23$ and normalisation 
$f_r = 0.14 \pm 0.02$. The best fitting bias and effective halo mass are almost 
identical to those determined with the simple luminosity limit HOD, with 
$b_g$ = 2.93 $\pm$ 0.19 and $M_{eff}$ = 10.3 $\pm$ 1.9 $\times10^{13} h^{-1}M_{\sun}$. 
This is reassuring, since it implies that the determination 
of the typical halo mass is largely independent of the form of the HOD, providing that 
it yields a reasonable fit to the clustering and space density. 

We also show in the bottom panel of Figure \ref{fig:NMfrac} $F_{\rm rad}(M)$ determined 
by dividing the best--fitting luminosity limit HODs to the radio and matched samples 
(shown by the blue and red lines in Figure \ref{fig:NM}). 
While both HOD fits, with and without the power law, provide good statistical fits 
to our data, they diverge for halo masses above $10^{15} h^{-1}M_{\sun}$ (where the power--law 
fit suggests that $\sim$60\% of central cluster galaxies would be radio--loud, while the 
luminosity--limit HOD suggests that the fraction levels off at $\sim30$\%) and below  
a few times $10^{13} h^{-1}M_{\sun}$ (where the luminosity--limit HOD suggests that  
there should be a sharp cut off in the fraction of radio--loud LRGs). 
Since the current 2SLAQ LRG data sample contains very few objects with halo masses 
above $10^{15} h^{-1}M_{\sun}$ or below $3\times10^{13} h^{-1}M_{\sun}$, a larger sample of radio 
galaxies is needed to test between different models in this regime.  

\section{Discussion and Conclusions}
\label{sec:disc}
We have measured the two--point cross-correlation function of a sample of radio--detected 
LRGs at $z\simeq$ 0.55 and a sample of radio--quiet LRGs that matches the luminosity, 
colour and redshift distribution of the radio loud sample. 
We find that radio--detected LRGs at $z\simeq$ 0.55 are significantly more clustered 
than the matched radio--quiet sample, with clustering scale lengths ($r_0$) of 
12.3$\pm$1.2 and 9.02$\pm$0.52 $h^{-1}$ Mpc respectively. 
This result suggests that the radio--detected LRGs typically occupy more massive haloes 
than other LRGs of the same optical luminosity and stellar mass. We confirm this by 
fitting HODs to the observed clustering, and show that the 
radio--detected LRGs have a typical halo mass of 10.1$\pm$1.4 $\times 10^{13} h^{-1}M_{\odot}$ 
and a bias of 2.96$\pm$0.17, compared to a halo mass of 6.44$\pm0.32\times 10^{13} h^{-1}M_{\odot}$ 
and a bias of 2.49$\pm$0.02 for the radio--quiet LRGs. 

The clustering of the radio--detected LRGs is best fitted by a HOD containing only 
galaxies which occupy the centres of haloes and are not satellites, though the current 
data do not allow us to exclude models with some radio satellite galaxies. 
We show that the dependence on radio fraction on halo mass can be modelled as a power-law 
with slope 0.65 $\pm$ 0.23, although it will take a larger samples of radio galaxies 
with a more precise clustering measurement to rule out other forms of this dependence.
Such samples are available at lower redshift, such as the SDSS/NVSS/FIRST sample presented 
by Best et al. (2005), which would also allow a measurement of any evolution in this relationship.
Mandelbaum et al. (2008) use both a clustering and weak lensing analysis of this sample to show that the 
the halo masses of radio loud galaxies are about twice the mass of a comparable radio quite sample, in 
good agreement with the results presented here.  

\begin{table}
  \begin{center}
    \caption{\label{tab:detrate} Radio--detected fraction of brightest--cluster galaxies 
    (above a limiting 1.4\,GHz radio power P$_{1.4}$) in the local
    universe and at $z\sim0.55$.  }
    \begin{tabular}{lrrl} 
      \multicolumn{1}{c}{Sample} & \multicolumn{2}{c}{P$_{1.4}$ (W\,Hz$^{-1}$)} & 
      \multicolumn{1}{c}{Reference} \\
      &  \multicolumn{1}{c}{$\geq10^{23}$} & \multicolumn{1}{c}{$\geq10^{24}$} \\
\hline \hline 
	BCGs, $z<0.11 $ & 33.3\%     & 22.2\%     & Burns 1990 \\
	BCGs, $z<0.2 $  & 32.7\%     & 19.9\%     & Lin \& Mohr 2007 \\
        BCGs, SDSS      & $\sim30$\%     & $\sim20$\% & Best et al.\ 2007 \\
\hline
             2SLAQ LRGs & ...        & $>30$\% & This paper \\
     \hline
    \end{tabular}
  \end{center}
\end{table}

Table \ref{tab:detrate} compares the radio detection rates for the most optically--luminous 
brightest cluster galaxies (BCGs) in the local universe with the 2SLAQ LRG value derived in 
\S5.4. The values found in the three local studies are remarkably consistent, and imply 
that the radio power above which at least 30\% of BCGs are detected rises from about 
$10^{23}$\,W\,Hz$^{-1}$ at $z\sim0.1$ to at least $10^{24}$\,W\,Hz$^{-1}$ at $z\sim0.55$. 

As noted by Johnston et al.\ (2008) low--power ($<10^{26}$\,W\,Hz$^{-1}$) radio galaxies in 
the 2SLAQ LRG sample have stellar populations which are generally no different from those in a 
radio--quiet comparison sample, and the majority of these radio galaxies have weak or no emission lines 
in their optical spectra.  These properties are consistent with a population of `low-excitation' 
radio galaxies powered by Bondi accretion of hot, X--ray emitting gas from the intergalactic medium 
(Allen et al.\ 2006; Hardcastle, Evans \& Croston 2007).  If this is correct, then the 
highly--clustered environment of the 2SLAQ radio galaxies is not surprising.  
The radio emission from central cluster galaxies could be 
be enhanced by the presence of a denser intracluster medium (ICM) 
which would both confine the radio lobes and provide a more 
efficient working surface for the radio jets, thus boosting the
observed radio luminosity (Barthel \& Arnaud 1996; Allen et
al.\ 2006).  However, Best et al.\ (2006) argue that this
boosting of radio emission from cluster galaxies only applies
to powerful radio sources which extend well beyond the host
galaxy and into the intracluster medium.  Since most of the
radio--detected 2SLAQ LRGs are low-power radio galaxies
whose radio emission is generally compact and confined
within the host galaxy, this ICM-related boosting may not 
play a strong role.

The most likely alternative is that the radio emission from
central cluster galaxies is boosted because of a higher fuelling
rate of gas onto the central black hole. This is plausible 
because gas from both the galaxy's own X--ray halo and a larger-scale
cluster cooling flow can contribute to the fuelling of an active nucleus 
in central cluster galaxies (Best et al.\ 2007).

\section*{Acknowledgements}

We acknowledge the support of the Australian Research Council through the award 
of an ARC Australian Professorial Fellowship to EMS and a QEII Fellowship to SMC. 
This work was also supported in part by a rolling grant from the UK STFC.  

The authors would like to thank Ravi Sheth for assistance with implementing the 
cross-correlation function calculation in the halo model, and Richard Bower, 
Shaun Cole, Russell Smith and the anonymous referee for helpful discussions and 
insightful comments on this work.
The authors thank the 2SLAQ team and the AAO staff for their contributions to 
the collection of the 2SLAQ data (http://www.2slaq.info/). DAW thanks the 
Department of Astronomy at the University of Illinois for their regular hospitality.

We would like to thank Cameron McBride, Jeff Gardner, and Andy Connolly for 
providing a pre-release version of the Ntropy package.  
Ntropy was funded by the NASA Advanced Information Systems Research Program grant NNG05GA60G.

Funding for the SDSS and SDSS-II has been provided by the Alfred P.
Sloan Foundation, the Participating Institutions, the National Science
Foundation, the U.S. Department of Energy, the National Aeronautics
and Space Administration, the Japanese Monbukagakusho, the Max Planck
Society, and the Higher Education Funding Council for England. The
SDSS Web Site is http://www.sdss.org/.

The SDSS is managed by the Astrophysical Research Consortium for the
Participating Institutions. The Participating Institutions are the
American Museum of Natural History, Astrophysical Institute Potsdam,
University of Basel, Cambridge University, Case Western Reserve
University, University of Chicago, Drexel University, Fermilab, the
Institute for Advanced Study, the Japan Participation Group, Johns
Hopkins University, the Joint Institute for Nuclear Astrophysics, the
Kavli Institute for Particle Astrophysics and Cosmology, the Korean
Scientist Group, the Chinese Academy of Sciences (LAMOST), Los Alamos
National Laboratory, the Max-Planck-Institute for Astronomy (MPIA), the
Max-Planck-Institute for Astrophysics (MPA), New Mexico State
University, Ohio State University, University of Pittsburgh,
University of Portsmouth, Princeton University, the United States
Naval Observatory, and the University of Washington.

\appendix
\section{Fitting the halo model to a magnitude limited sample}

\begin{figure*}
\vspace{7.5cm}
\includegraphics{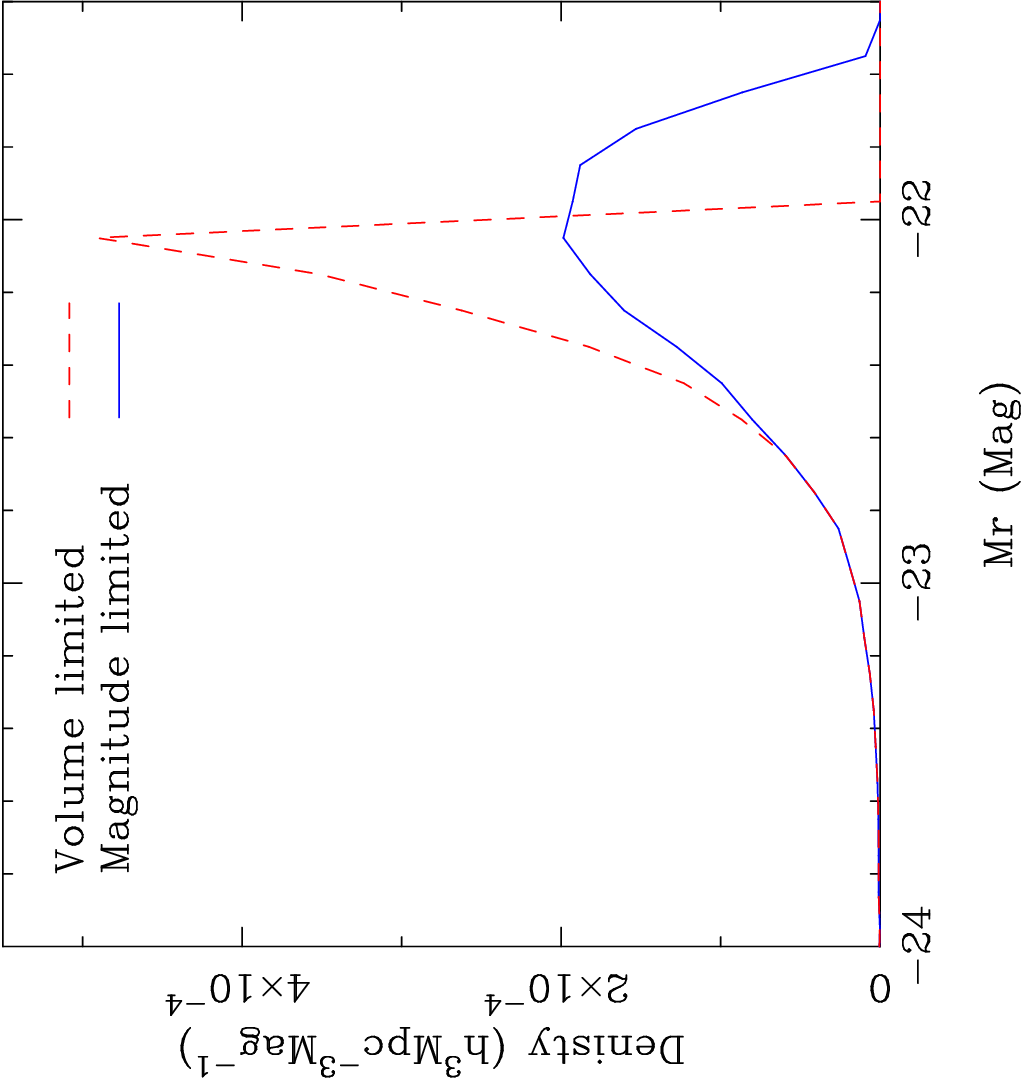}
\includegraphics{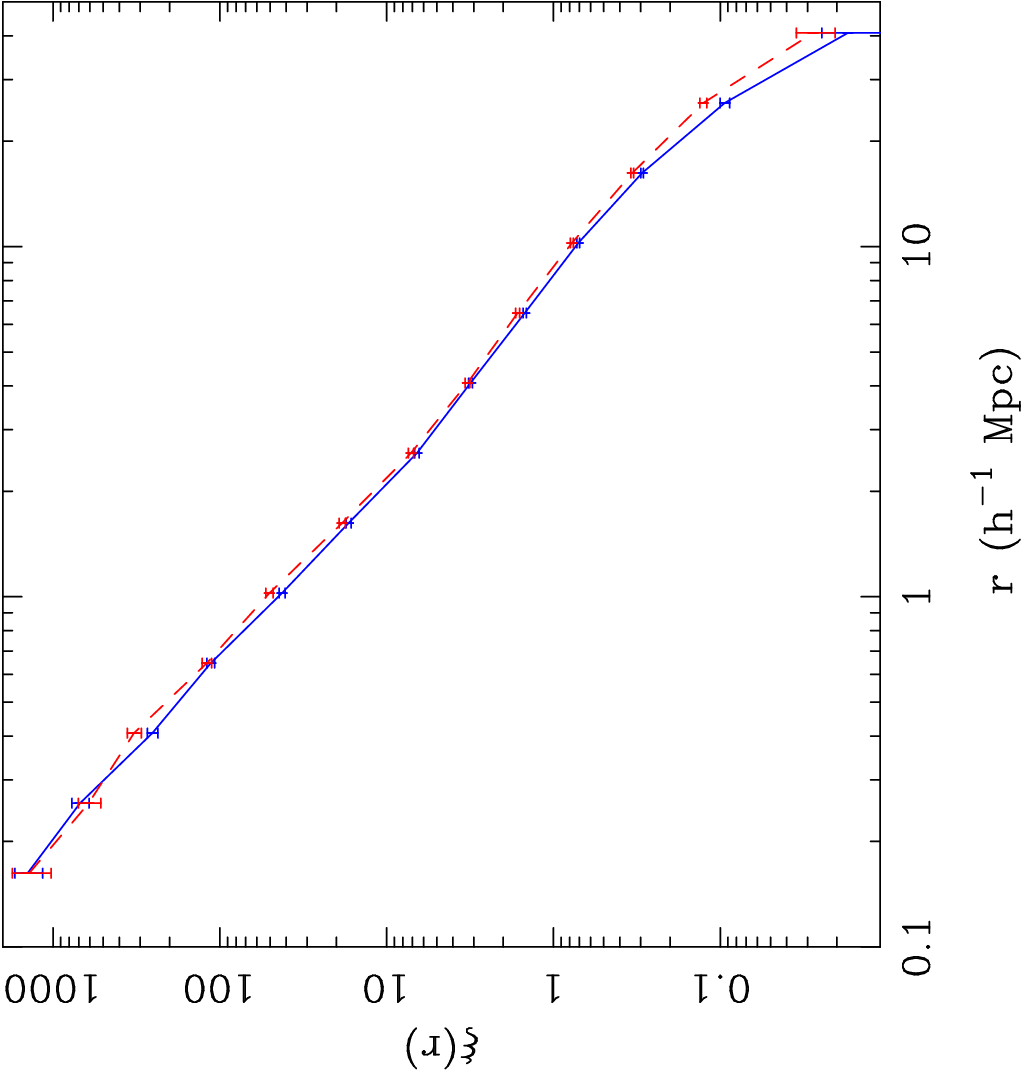}

\caption{\label{fig:galform} The absolute magnitude distribution (left) and 
the real space 2pt--correlation function (right) for the volume limited (dashed line) 
and magnitude--limited (solid line) simulated galaxy samples.}
\end{figure*}

In section \ref{sec:halo} we estimate the bias and effective halo mass of effectively 
magnitude--limited galaxy samples using the halo model framework. The halo occupation 
distribution we use is designed to apply to volume limited galaxy samples and so its application here may 
result in biased estimates of these parameters. In this appendix we use the millennium simulation 
to investigate the magnitude of this bias and show that it is likely to be only $\sim$5\%, 
and comparable to the likely systematic error introduced by any differences between 
the assumed form of the HOD and the actual form.

\begin{figure}
\vspace{7.5cm}
\includegraphics{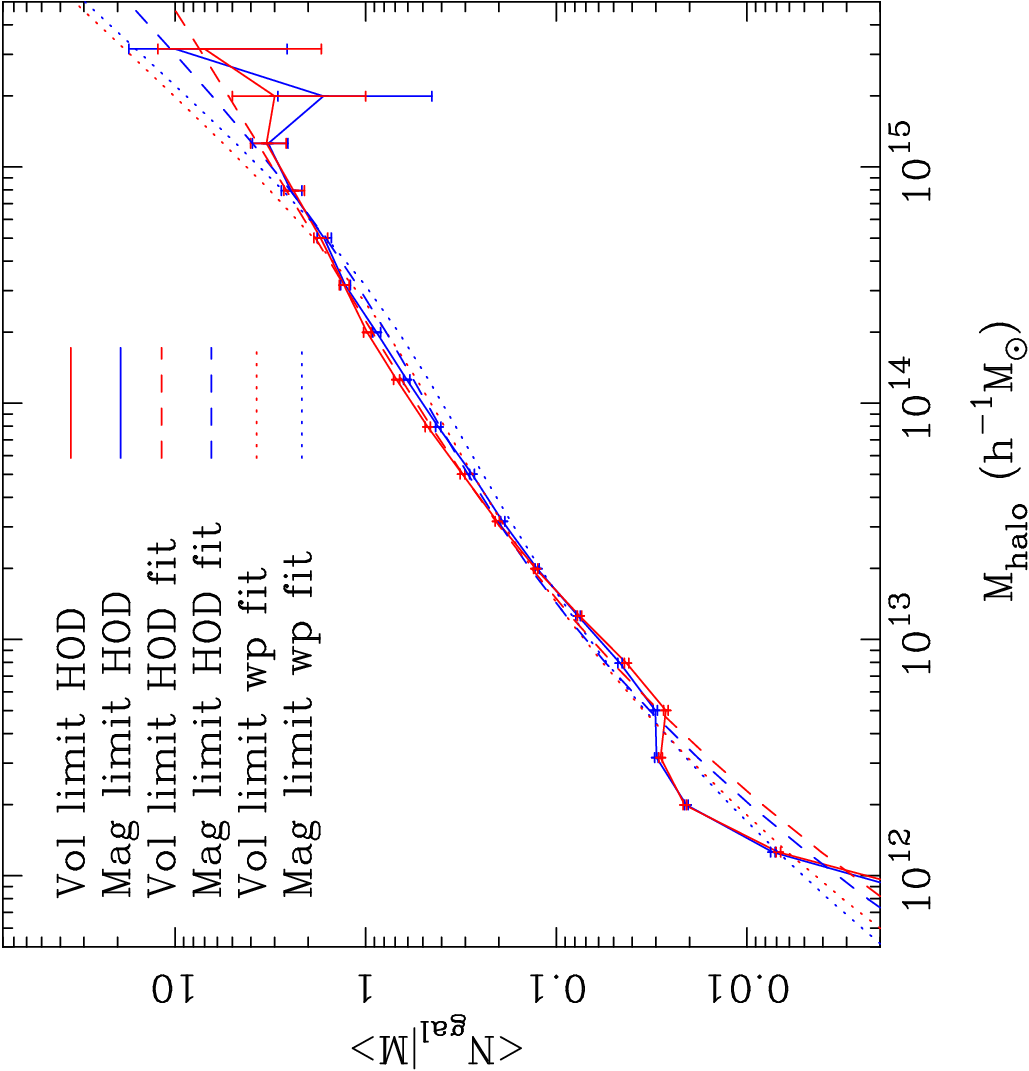}

\caption{\label{fig:galformHOD} The measured HOD for the volume (red) and magnitude (blue) limited samples, as well as analytic HOD fits to the measured HODs (dashed lines) and the clustering (dotted lines).}
\end{figure}

We construct two samples of galaxies generated using the latest version of the Durham group semi-analytic galaxy formation model detailed in Font et al. (2008). For the first sample we simply select all galaxies in the z = 0.02 volume with SDSS $r$ magnitudes $< -22$ i.e. a volume limited sample of galaxies with a space density equivalent to our full LRG sample. We also wish to generate a magnitude limited sample, and so calculate the apparent SDSS r magnitude of each galaxy in our volume assuming an observer at one edge of that volume. We then limit this sample to $r < 19.67$ which matches the space density of the volume limited sample. Both samples contain $\sim20,000$ galaxies. Fig. \ref{fig:galform} shows the absolute magnitude distribution and real space 2pt-auto correlation function for these samples. As one would expect the magnitude limited sample which contains some intrinsically fainter galaxies shows a slightly lower clustering amplitude. In order to test whether we are able to accurately measure the effective halo mass by fitting a HOD to the clustering measurements we first determine the actual HOD of the two galaxy samples shown as the solid lines in Fig. \ref{fig:galformHOD}. We then take a suitable analytic form of the HOD and fit it to both the actual HOD from the simulation and the derived 2pt--correlation function and space density, shown as the dashed and dotted lines respectively. It is worth noting that the actual HOD derived from GALFORM does not exactly match our analytic form at low masses and would not match any form typically used in the literature. The clear 'bump' at $10^{12} < M_{halo} < 10^{13}$ is caused by the onset of AGN feedback in model (Carlton Baugh, Private Communication). It will be interesting to see if there is any evidence for this in directly determined HODs of massive galaxies. 

\begin{table} 
  \begin{center}
    \caption{\label{tab:halomass}The Effective halo mass in units of $10^{13}h^{-1}M_{\sun}$.}
    \begin{tabular}{c  c  c} 
      \multicolumn{1}{c}{Method} &
      \multicolumn{1}{c}{Volume Limited} &
      \multicolumn{1}{c}{Magnitude Limited}\\
      \hline \hline 
	Actual         & 6.63  &  6.38\\
	HOD fit        & 6.98 $\pm$ 0.09  &  6.87 $\pm$ 0.08\\
	Clustering fit & 7.01 $\pm$ 0.11 &  6.78 $\pm$ 0.11\\
      \hline
    \end{tabular}
  \end{center}
\end{table}

Despite the differences in the analytic and actual HODs the analytic HOD does 
provide good fits to the clustering and space density of the samples  within the errors. 
Table \ref{tab:halomass} gives the values $M_{eff}$ 
determined from fitting the analytic HOD to the measured HOD and the clustering 
as well as that determined directly from the simulation. $M_{eff}$ is overestimated 
by $\sim$7\% fitting the analytic form to either measured HOD or clustering for both 
the volume and magnitude limited samples, due to the excess of galaxies in the `bump' 
in the HOD at low halo masses. The difference between the fits to the HOD and the 
clustering is larger for the magnitude--limited sample, but is still only at the 2\% 
level and is consistent with the error on the fit to the clustering. 
We can therefore be confident that the likely magnitude of any systematic error 
on $M_{eff}$ caused by fitting the HOD to a magnitude limited rather than volume--limited 
sample is less than the measurement errors for all the samples considered in Section \ref{sec:halo}. 
It is interesting to note that a larger systematic error in $M_{eff}$ would be introduced if 
the true HOD is not well represented by the assumed analytic form of the HOD. However, one 
would expect still relative measurements of $M_{eff}$ to be representative even in this case.    

\bsp
\label{lastpage}


\begin{thebibliography}{99}
\bibitem{} Allen, S.W., Dunn, R.J.H., Fabian, A.C., Taylor, G.B., Reynolds, C.S., 2006, MNRAS, 372, 21
\bibitem{} Auriemma, C., Perola, G.C., Ekers, R.D., Fanti, R., Lari, 
   C., Jaffe, W.J., Ulrich, M.H., 1977, A\&A, 57, 41 
\bibitem{} Barthel, P., Arnaud, K., 1996, MNRAS. 284, L45
\bibitem{} Becker, R.H., White, R.L., Helfand, D.J., 1995, ApJ, 450, 559 
\bibitem{} Berlind A.~A., Weinberg D.~H., 2002, ApJ, 575, 587
\bibitem{} Best, P.N., Kauffmann, G., Heckman, T, Brinchmann, J., Charlot, S.,Ivezic, Z., 
   White, S.D.M., 2005a, MNRAS 362, 25 
\bibitem{} Best, P.N., Kauffmann, G., Heckman, T, Ivezic, Z., 2005b, MNRAS 362, 9
\bibitem{} Best, P.N., Kaiser, C.R., Heckman, T.M., Kauffmann, G., 2006, MNRAS, 368, L67
\bibitem{} Best, P.N., von der Linden, A., Kauffmann, G., Heckman, T.M., Kaiser, C.R., 2007, MNRAS, 894, 908 
\bibitem{} Blake, C., Wall, J., 2002, MNRAS, 337, 993 
\bibitem{} Blake, C., Mauch, T., Sadler, E., 2004, MNRAS, 347, 787 
\bibitem{} Bock, D. C.-J.,  Large, M. I., Sadler, E.M., 1999, AJ, 117, 1578
\bibitem{} Bower R.~G., Benson A.~J., Malbon R., Helly J.~C., Frenk C.~S., Baugh 
C.~M., Cole S., Lacey C.~G., 2006, MNRAS, 370, 645
\bibitem{} Cannon, R.D. et al., 2006, MNRAS, 372, 425 
\bibitem{} Ciotti L., Ostriker J.~P., 2007, ApJ, 665, 1038 
\bibitem{} Colless, M. et al., 2001, MNRAS, 328, 1039 
\bibitem{} Condon, J.J., Cotton, W.D., Greisen, E.W., Yin, Q.F., Perley, R.A., Taylor, G.B., 
   Broderick, J.J., 1998, AJ, 115, 1693 
\bibitem{} Cooray A., Sheth R., 2002, PhR, 372, 1
\bibitem{} Croton D.~J., et al., 2006, MNRAS, 365, 11
\bibitem{} Davis M., Peebles P.~J.~E., 1983, ApJ, 267, 465
\bibitem{} Fanaroff, B,L, Riley, J.M., 1974, MNRAS, 167, 31 
\bibitem{} Font A.~S., et al., 2008, MNRAS, 389, 1619
\bibitem{} Gardner, J.~P., Connolly, A., \& McBride, C.\ 2007, ArXiv e-prints, 709, arXiv:0709.1967
\bibitem{} Hardcastle, M.J., Evans, D.A., Croston, J.H., 2007, MNRAS, 376, 1849 
\bibitem{} Heckman, T.M., Smith, E.P., Baum, S.A., van Breugel, W.J.M., Miley, G.K., 
Illingworth, G.D., Bothun, G.D., Balick, B., 1986, ApJ, 311, 526 
\bibitem{} Jing Y.~P., Mo H.~J., Boerner G., 1998, ApJ, 494, 1 
\bibitem{} Johnston, H.M., Sadler, E.M., Cannon, R., Croom, S.M., Ross, N.P., Schneider, D.P., 
2008, MNRAS, 384, 692 
\bibitem{} Ledlow, M.J., Owen, F.R., 1996, AJ, 112, 9
\bibitem{} Ma C.-P., Fry J.~N., 2000, ApJ, 543, 503 
\bibitem{} Mandelbaum R., Li C., Kauffmann G., White S.~D.~M., 2008, arXiv, arXiv:0806.4089
\bibitem{} Mauch, T., Sadler, E.M., 2007, MNRAS, 375, 931
\bibitem{} Navarro, J.~F., Frenk, C.~S., \& White, S.~D.~M.\ 1996, ApJ, 462, 563
\bibitem{} Overzier, R. A., R\"ottgering, H.J.A., Rengelink, R.B., Wilman, R.J., 2003, A\&A, 
405, 530 
\bibitem{} Parma P., Murgia M., Morganti R., Capetti A., de Ruiter H.~R., Fanti R., 1999, A\&A, 344, 7 
\bibitem{} Peacock J.~A., Smith R.~E., 2000, MNRAS, 318, 1144 
\bibitem{} Prestage, R.M., Peacock, J.A., 1988, MNRAS, 230, 131 
\bibitem{} Sadler, E.M. et al., 2007, MNRAS, 381, 211 
\bibitem{} Scoccimarro R., Sheth R.~K., Hui L., Jain B., 2001, ApJ, 546, 20
\bibitem{} Scranton, R., et al.\ 2002, ApJ, 579, 48
\bibitem{} Seljak U., 2000, MNRAS, 318, 203
\bibitem{} Sheth, R.~K., \& Tormen, G.\ 1999, MNRAS, 308, 119
\bibitem{} Springel V., et al., 2005, Natur, 435, 629
\bibitem{} Wake D.~A., et al., 2006, MNRAS, 372, 537
\bibitem{} Wake D.~A., et al., 2008, MNRAS, 387, 1045
\bibitem{} York, D.G. et al., 2000, AJ 120, 1579
\bibitem{} Zehavi, I., et al.\ 2005, ApJ, 621, 22

\end{thebibliography}
\end{document}